\documentclass[a4paper,11pt]{article}
\usepackage{pos}

\title{$B$ Physics: From Present to Future Colliders}
\author*[a,b]{Monika Blanke}
\affiliation[a]{Institut f\"ur Astroteilchenphysik, Karlsruhe Institute of Technology,\\
  Hermann-von-Helmholtz-Platz 1,
  D-76344 Eggenstein-Leopoldshafen, Germany}

\affiliation[b]{Institut f\"ur Theoretische Teilchenphysik,
  Karlsruhe Institute of Technology, \\
Engesserstra\ss e 7,
  D-76128 Karlsruhe, Germany}

\emailAdd{monika.blanke@kit.edu}
\abstract{These proceedings provide a brief overview of the status of $B$ meson physics, putting particular emphasis on precision tests of the Standard Model with meson mixing data, and on the anomalies in charged- and neutral-current semileptonic $B$ decays. In addition to summarising the current status,  some promising directions to be pursued at future collider experiments are highlighted.}

\FullConference{Corfu Summer Institute 2023 ``School and Workshops on Elementary Particle Physics and Gravity'' (CORFU2023)\\
 23 April -- 6 May  and 27 August -- 1 October, 2023\\
Corfu, Greece\\}

\begin{document}
\maketitle

\section{Introduction}

One of the big mysteries of the Standard Model (SM) is the flavour structure of its matter content.
It is generated by the Yukawa coupling matrices which exhibit a very hierarchical pattern, observable in the quark and lepton masses spanning several orders of magnitude. In addition, the quark Yukawa couplings are nearly aligned with each other, leading to a highly non-generic structure of the CKM matrix. The origin of these hierarchies  is unknown and commonly referred to as the SM flavour puzzle. 

Besides calling for an extension of the SM to explain the origin of flavour and its hierarchies, this peculiarity of the SM also leads to a strong suppression of flavour-changing neutral currents (FCNCs), leaving ample room for potential New Physics (NP) contributions. FCNCs are further suppressed in the SM by the unitarity of the CKM matrix, leading to the absence of tree-level contributions and the suppression of loop contributions via the GIM-mechanism. Lastly, also the chiral structure of the SM weak interactions contributes to the smallness of FCNCs\, e.\,g.\ in leptonic decays of pseudoscalar mesons such as $B_s\to \mu^+\mu^-$. 

FCNCs therefore offer one of the most promising tools to probe
 new particles and interactions beyond the SM, even if they are too heavy or too weakly coupled to be directly observable in high-energy collisions at the Large Hadron Collider (LHC).
 In order to maximise the impact of this endeavour, high precision is required both in the experimental measurements of flavour-violating decays and in their theory predictions.

\section{Status of CKM determinations and meson mixings}

One crucial ingredient of precise SM predictions for flavour-violating decays is the determination of the parameters of the CKM matrix with high accuracy. In order to achieve a clean measurement of CKM parameters free from potential NP contributions, the traditional approach is to determine them from charged-current decays, These processes arise from tree-level $W$-boson exchanges in the SM and are therefore expected to be insensitive to NP effects.

Due to the CKM unitarity, four independent parameters determine the entire CKM matrix. Tree-level charged-current decays can be used to measure the off-diagonal elements $|V_{us}|$, $|V_{ub}|$, and $|V_{cb}|$. Unfortunately, at the moment all of them are subject to tensions in the data \cite{ParticleDataGroup:2022pth}. $|V_{us}|$ values extracted from leptonic and semileptonic kaon decays, and from $\tau$-lepton decays to strange particles are in some disagreement with each other. In addition, tests of first-row unitarity exhibit a deviation at the $3\sigma$ level from the prediction $|V_{ud}|^2+|V_{us}|^2 = 1$, known as the Cabibbo angle anomaly \cite{Crivellin:2022rhw}. $|V_{ub}|$ and $|V_{cb}|$ can both be measured in inclusive and exclusive semileptonic $B$ meson decays. For both CKM elements, long-standing tensions between the two kinds of determinations exist, leading to significant uncertainties in their values \cite{Gambino:2020jvv}. In order to fully profit from future experimental improvements, a better control of the underlying theoretical uncertainties is hence badly needed.

The fourth CKM parameter accessible in tree-level decays is the angle $\gamma$ $(\phi_3)$ of the Unitarity Triangle. Its measurement in $B\to D K$ decays is theoretically extremely clean \cite{Brod:2013sga}, and the experimental precision is currently limited by statistics. Future measurements at LHCb and Belle~II are therefore expected to improve the knowledge of $\gamma$ to degree-level precision \cite{Belle-II:2018jsg,LHCb:2012myk}.

\begin{figure}
\centering{\includegraphics[width=.65\textwidth]{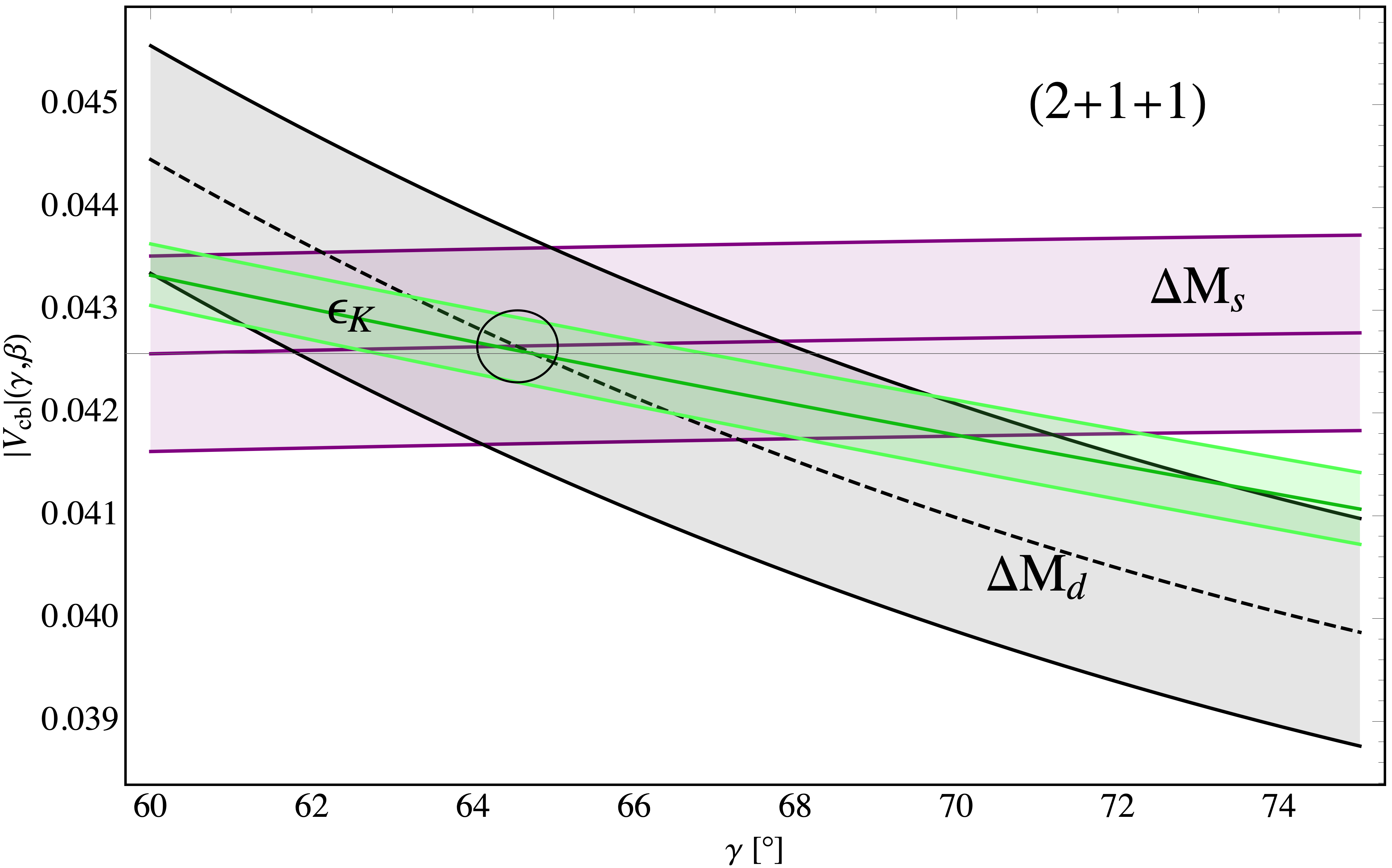}}
\caption{\label{fig:DelF2}CKM determination from $\Delta F = 2$ measurements, using 2+1+1 flavour lattice QCD data. Figure taken from~\cite{Buras:2022wpw}.}
\end{figure}

The current lack of precision in CKM determinations from tree-level decays has led to the suggestion of alternative strategies. Notably, significantly higher accuracies can be reached in neutral meson mixing observables. While previous determinations of the relevant hadronic matrix elements using 2+1 flavour lattice QCD \cite{FermilabLattice:2016ipl} hinted at some inconsistencies in the SM description of $\Delta F=2$ data 
\cite{Blanke:2016bhf,Blanke:2018cya,King:2019rvk}, using the more recent 2+1+1 flavour results from lattice QCD \cite{Dowdall:2019bea} a perfectly consistent SM description of the data is obtained \cite{Blanke:2019pek,Buras:2022wpw}, see Fig.\ \ref{fig:DelF2}. While this does not strictly exclude hidden NP effects in meson mixing observables, it suggests that $\Delta F =2$ processes are governed by SM physics. Consequently they can be used to determine the elements of the CKM matrix with a precision that by far surpasses the one obtained from tree-level decays. Note that a similar approach is taken by using the results of global CKM fits \cite{UTfit:2022hsi,Charles:2020dfl} which combine the available information from tree-level and $\Delta F =2$ processes.

 The results of this exercise can then be used to make precise SM predictions for rare $K$ and $B$ decays that can in turn be tested against experiment \cite{Buras:2022qip}. In the case of kaon decays, the current experimental precision is not sufficient to make practical use of such high accuracy in the SM predictions. Substantial improvements could however be achieved in future dedicated kaon experiments. In $B$ decay observables, on the other hand, significant tensions can be identified in $b\to s\mu^+\mu^-$ transitions, currently driven mainly by LHCb data. Clearly, an independent experimental confirmation, e.\,g.\ by Belle II, as well as a better theory understanding of hadronic uncertainties is required before being able to draw definite conclusions about the presence of NP.

\section{Persisting flavour anomalies}

The previously mentioned tensions in $b\to s\mu^+\mu^-$ observables are among the flavour anomalies that attract a lot of attention within the flavour community and beyond. While the directly related deviations from lepton flavour universality in the ratios $\mathcal{R}(K)$ and $\mathcal{R}(K^*)$ have found an experimental resolution, the tensions in branching ratios and angular observables of $b\to s\mu^+\mu^-$ transitions persist.

While the hints for lepton flavour universality violation in $\mathcal{R}(K^{(*)})$, testing the first two lepton generations, have vanished, the persisting anomaly in the ratios $\mathcal{R}(D^{(*)})$ still indicates a potential violation of lepton universality for the third lepton generation. In what follows we first review the status of the $\mathcal{R}(D^{(*)})$ anomaly and discuss possible probes in current and future experiments. We then turn to the situation in $b\to s \mu^+\mu^-$ transitions, where again we complete the presentation with an outlook to possible tests at future colliders.

\subsection{Semileptonic charged-current anomalies: $\mathcal{R}(D^{(*)})$}

Lepton flavour universality in semileptonic $B\to D^{(*)}$ decays can conveniently be probed in the ratios
\begin{equation}
\mathcal{R}(D^{(*)})=\frac{\text{BR}(B\to D^{(*)}\tau\nu)}{\text{BR}(B\to D^{(*)}\ell\nu)}\,,\qquad \ell=e,\mu
\end{equation}
comparing the decay rates into third-generation leptons with those into first- and second-generation leptons. Note that due to the mass of the $\tau$-lepton, these ratios deviate from one already in the SM. For the same reason, the cancellation of hadronic uncertainties  in these ratios is not complete, yet they are still significantly cleaner than the individual branching ratios. Contrary to the latter, they are also independent of the size of $|V_{cb}|$.

Multiple measurements of $\mathcal{R}(D^{(*)})$ by various experimental collaborations exist \cite{BaBar:2012obs,BaBar:2013mob,Belle:2015qfa,Belle:2016ure,Belle:2016dyj,Belle:2017ilt, Belle:2019rba,LHCb:2015gmp,LHCb:2017smo,LHCb:2017rln,LHCb:2023zxo,Belle-II:2024ami}, leading to a world average which is $3.3\sigma$ above the SM prediction \cite{HFLAV:2022esi}. Notably, this anomaly has been with us for more than a decade. Additionally, the LHCb collaboration found the analogous ratio $\mathcal{R}(J/\psi)$ to be surprisingly large \cite{LHCb:2017vlu}, however in this case the corresponding SM value is less precisely known. 
Note that while recent form factor determinations using lattice QCD data significantly ameliorate the tension in $\mathcal{R}(D^{(*)})$ \cite{DiCarlo:2021dzg,Martinelli:2021frl,FermilabLattice:2021cdg}, they create inconsistencies in the longitudinal polarisations $F_L^{e,\mu}$ which can not be resolved even in the presence of NP \cite{Fedele:2023ewe}.

Another interesting observable testing lepton flavour universality in semileptonic $b \to c $ transitions is the ratio of baryonic decay rates
\begin{equation}
\mathcal{R}(\Lambda_c)=\frac{\text{BR}(\Lambda_b\to \Lambda_c\tau\nu)}{\text{BR}(\Lambda_b\to \Lambda_c\ell\nu)}\,,\qquad \ell=e,\mu\,.
\end{equation}
Its first measurement by the LHCb collaboration \cite{LHCb:2022piu} turned out to be somewhat lower than the SM prediction, albeit still consistent due to sizeable uncertainties. However it has been shown that the observables $\mathcal{R}(D^{(*)})$ and $\mathcal{R}(\Lambda_c)$ are connected via a model-independent sum 
rule~\cite{Blanke:2018yud,Blanke:2019qrx,Fedele:2022iib}
\begin{equation}
\frac{\mathcal{R}(\Lambda_c)}{\mathcal{R}(\Lambda_c)_\text{SM}} = 0280 \frac{\mathcal{R}(D)}{\mathcal{R}(D)_\text{SM}} + 0.720 \frac{\mathcal{R}(D^*)}{\mathcal{R}(D^*)_\text{SM}}\,,
\end{equation}
so that an enhancement of $\mathcal{R}(D^{(*)})$ implies a predicted value of $\mathcal{R}(\Lambda_c)$ above the SM. Note that this prediction holds irrespective of the concrete underlying short-distance dynamics. While the sum rule has originally been derived assuming NP only in $b\to c\tau\nu$ transitions \cite{Blanke:2018yud,Blanke:2019qrx}, in a more recent analysis also NP in the light lepton modes has been included into the analysis \cite{Fedele:2022iib}. It turned out that NP in $b \to c\ell\nu$ large enough to significantly alter the sum rule prediction is ruled out by complementary constraints from  CKM fits, angular distribution and $D^*$ polarisation data as well as high-energy collider bounds. A future confirmation of a suppressed $\mathcal{R}(\Lambda_c)$ would thus be incompatible with the measured enhancement of $\mathcal{R}(D^{(*)})$, hinting at an underlying experimental issue.

Possible NP solutions to the $\mathcal{R}(D^{(*)})$ anomaly have been discussed at length in the literature, see e.\,g.\ \cite{Fischer:2021sqw,London:2021lfn,Crivellin:2022qcj} for recent reviews. Viable solutions include tree level contributions from a charged scalar boson (``charged Higgs''), scalar or vector leptoquarks. While a charged vector boson $W'$ could provide a good fit to the low energy $b\to c\tau\nu$ data, it is strongly disfavoured by  electroweak precision constraints and LHC searches for its $Z'$ partner. Interestingly, the charged Higgs solution is somewhat preferred by the data as it can explain the possible enhancement of the longitudinal $D^*$ polarisation $F_L^\tau$  \cite{Belle:2019ewo,LHCb:2023ssl}, while leptoquark models only have a minor impact on this observable.

A charged Higgs resolution of the $\mathcal{R}(D^{(*)})$ anomaly would imply sizeable rates for the decay $B_c \to\tau\nu$ \cite{Alonso:2016oyd}. While currently only weak indirect bounds on the corresponding branching ratio exist \cite{Blanke:2018yud,Aebischer:2021ilm}, a future $e^+e^-$ collider like FCC-ee can place stringent limits on this decay and thereby test charged Higgs effects underlying the anomaly \cite{Zuo:2023dzn}.

\begin{figure}
\centering{\includegraphics[width=.5\textwidth]{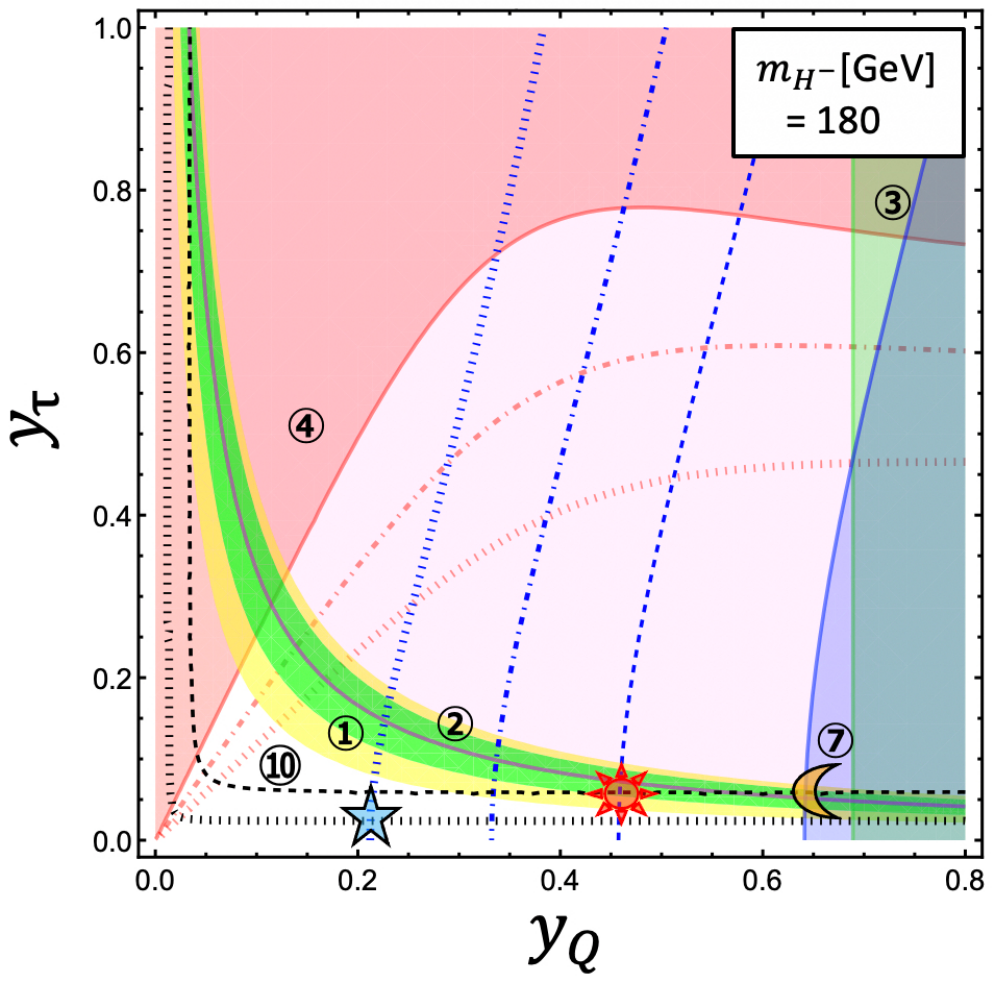}}
\caption{Current and projected future collider reach for the charged Higgs parameter space. The green and yellow bands display the parameter space predicted by the $\mathcal{R}(D^{(*)})$ anomaly. Figure taken from \cite{Blanke:2022pjy}.\label{fig:H}}
\end{figure}

Complementary information on the charged Higgs solution can be obtained from collider searches, in particular at the (HL-)LHC. For $m_{H^\pm}> 400\,\text{GeV}$ a resolution of the anomaly is excluded by $W'$ searches in the $\tau\nu$ final state, while for smaller masses this search becomes less effective due to the large $W\to\tau\nu$ 
background \cite{Iguro:2018fni}. Further constraints on the parameter space can be obtained by recasting limits from SUSY stau and (flavoured) dijet searches (red and blue contours in Fig.\ \ref{fig:H}) \cite{Iguro:2022uzz}. In addition, a substantially increased reach could be obtained by requiring an additional $b$-tagged jet in the final state of the $\tau\nu$ resonance search, due to the achieved strong background suppression \cite{Blanke:2022pjy}. Already with Run 2 LHC data the parameter space above the black dashed line  in Fig.\ \ref{fig:H} could be covered, while the HL-LHC phase would reach down to the dotted black line allowing to fully test the charged Higgs solution of the $\mathcal{R}(D^{(*)})$ anomaly.

\begin{figure}
\centering{\includegraphics[width=.39\textwidth]{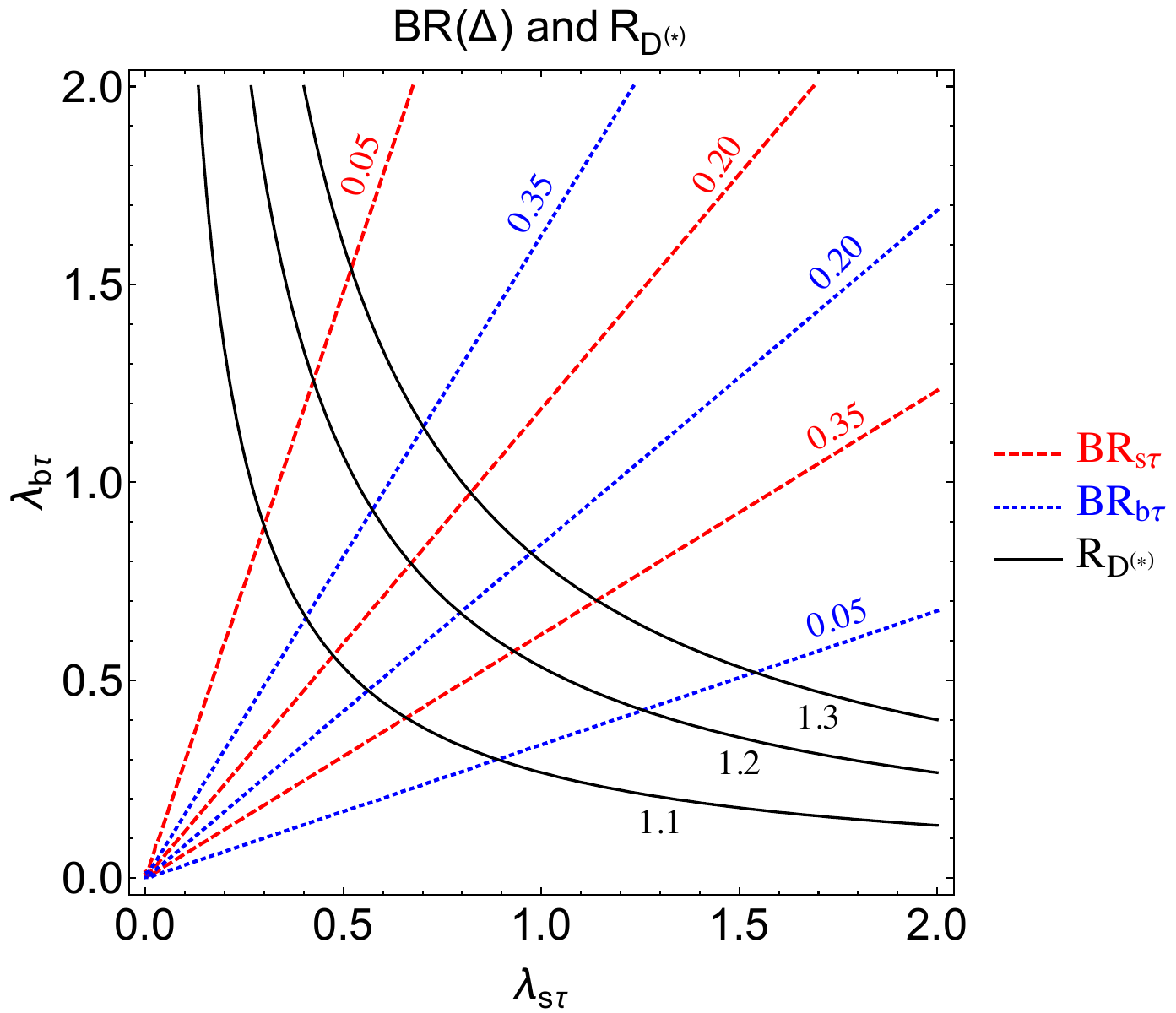}\includegraphics[width=.6\textwidth]{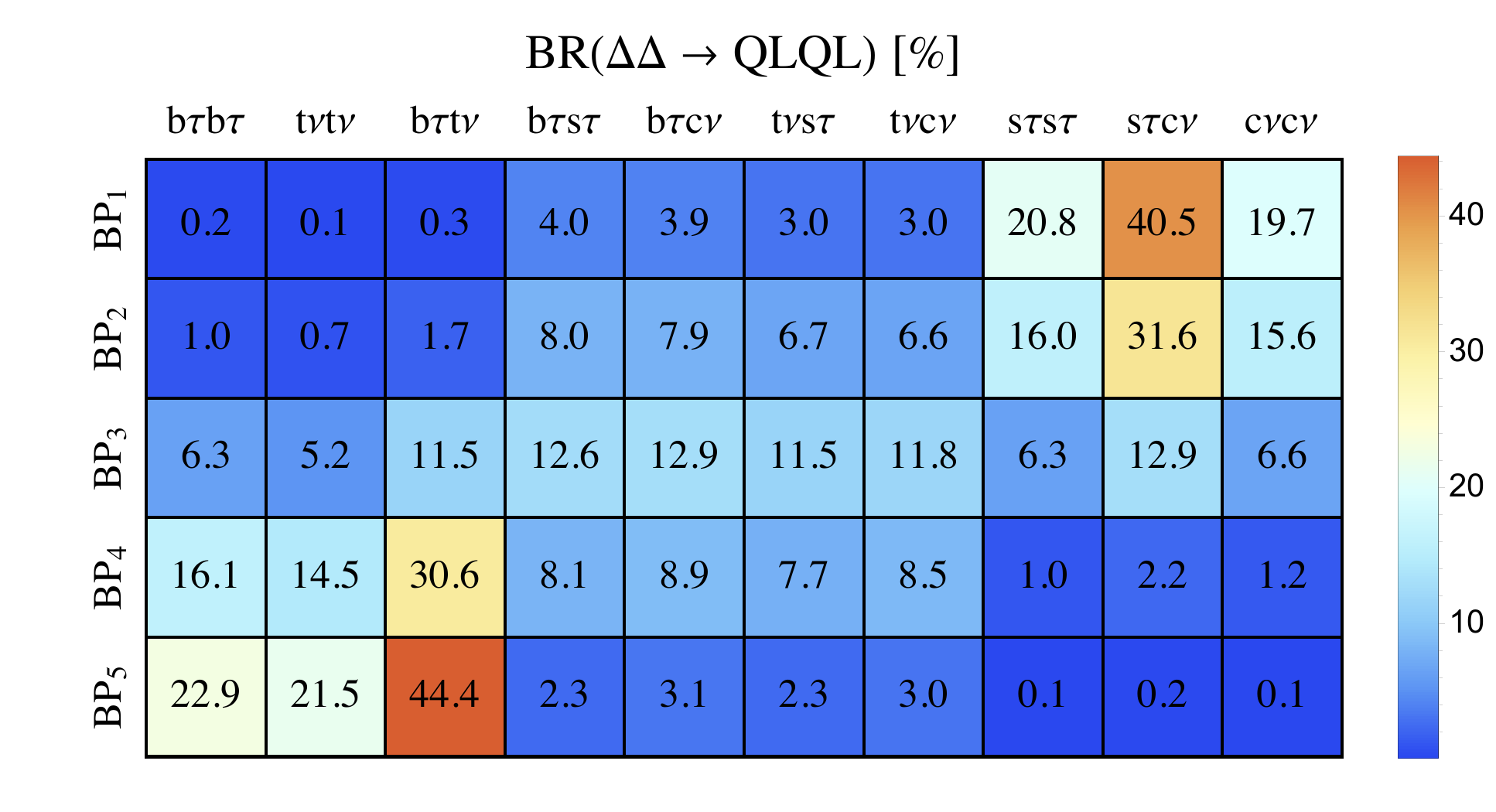}}
\caption{{\it Left:} Complementarity in $\mathcal{R}(D^{(*)})$ and leptoquark decay branching ratios in determining leptoquark couplings. {\it Right:} Branching ratios for leptoquark pairs for various benchmark points. Figures taken from \cite{Bernigaud:2021fwn}.\label{fig:LQ}}
\end{figure}

For leptoquark solutions of the anomaly, on the other hand, a promising avenue are pair production searches at the (HL-)LHC \cite{Diaz:2017lit}, as they are abundantly produced through QCD interactions. Beyond the ability to directly discover new particles, these searches also yield complementary information to the one obtained from the low-energy flavour observables. Measuring the branching ratios into multiple different final states allows to independently determine the relevant leptoquark coupling parameters and thus provide insight on the underlying flavour structure, while the $\mathcal{R}(D^{(*)})$ ratios are sensitive only to their product. This is shown in Fig.\ \ref{fig:LQ} for the case of the $SU(2)$-singlet vector leptoquark $\Delta$. Note that the mixed final state $b\tau\,t\nu$ is not only particularly sensitive to the relevant coupling ratios, but also experimentally rather distinct \cite{Bernigaud:2021fwn}.
 % Also the leptoquark mass scale

\subsection{Semileptonic neutral-current anomalies: $b\to s \mu^+\mu^-$}

Although the anomalies in the lepton flavour universality ratios $\mathcal{R}(K^{(*)})$ have vanished thanks to improvements in the experimental analysis \cite{LHCb:2022vje}, $B$ decays with the underlying semileptonic neutral-current transition $b\to s \mu^+\mu^-$ still exhibit a consistent pattern of deviations from their SM predictions. These are, on the one hand, various branching ratios like $\text{BR}(B\to K\mu^+\mu^-)$ \cite{LHCb:2014cxe} and $\text{BR}(B_s\to\phi\mu^+\mu^-)$~\cite{LHCb:2021zwz},  and the theoretically cleaner angular observables in $B\to K^*\mu^+\mu^-$ \cite{LHCb:2020gog}, on the other hand. For the latter, recent analyses have demonstrated that while part of the anomaly may be due to non-local SM contributions, the presence of new short-distance NP contributions is still preferred by the data \cite{Ciuchini:2022wbq,Hurth:2023jwr,LHCb:2023gel,Bordone:2024hui}. 

In fact, the global fits recently performed by various groups -- although based on partially different theoretical and experimental inputs -- show a good agreement with each other. They consistently show a significant pull towards large negative NP contributions to the Wilson coefficient $C_9$, parametrising the $(\bar bs)_{V-A}(\bar\mu\mu)_V$ operator. While, as mentioned above, non-local charm loop contributions could be responsible for part of the effect, a full resolution of the anomaly in terms of hadronic effects appears unlikely at present. It is worth noting that the recent SM-like findings for $\text{BR}(B_s\to\mu^+\mu^-)$ \cite{CMS:2022mgd,ATLAS:2018cur,LHCb:2021vsc} point towards the absence of significant NP contributions to  the axialvector current $(\bar bs)_{V-A}(\bar\mu\mu)_A$, entering the Wilson coefficient $C_{10}$. Notably, due to the SM-like ratios $\mathcal{R}(K^{(*)})$ such NP effects  would have to be lepton flavour universal and affect both $b\to s\mu^+\mu^-$ and $b\to s e^+e^-$ transitions. 

The NP scale behind this anomaly can be estimated in a model-independent way from the relevant SMEFT operator(s). Minimally, the anomaly requires NP effects in the operator\footnote{Note that a non-standard $C_9$ but SM-like $C_{10}$ would require NP also in the operator $(\bar Q_3 \gamma^\mu Q_2) (\bar E_2\gamma_\mu E_2)$, i.\,e.\ in right-handed leptons.}
\begin{equation}
\frac{1}{\Lambda^2}(\bar Q_3 \gamma^\mu Q_2) (\bar L_2\gamma_\mu L_2)\,,
\end{equation}
with $\Lambda \simeq 40\,\text{TeV}$. Clearly, such a large NP scale is beyond the reach of the (HL-)LHC and challenging even for a $100\,\text{TeV}$ proton collider. On the other hand, the $b\to s\mu^+\mu^-$ anomaly offers a promising physics case for a future multi-TeV muon collider where the anomaly could be directly accessed in the process $\mu^+\mu^-\to bs$ \cite{Huang:2021biu,Azatov:2022itm}. Lastly, due to the observed $e/\mu$ universality, also $e^+e^-$ colliders might in principle be suitable to test the anomaly, the required energy and luminosity remain however to be investigated.

\begin{figure}
\includegraphics[width=.49\textwidth]{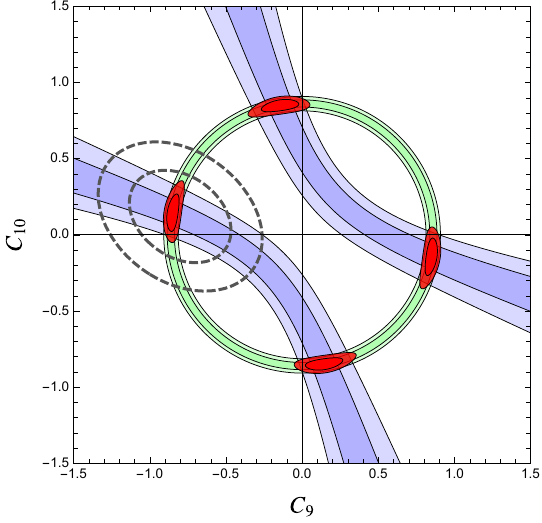}\hfill\includegraphics[width=.49\textwidth]{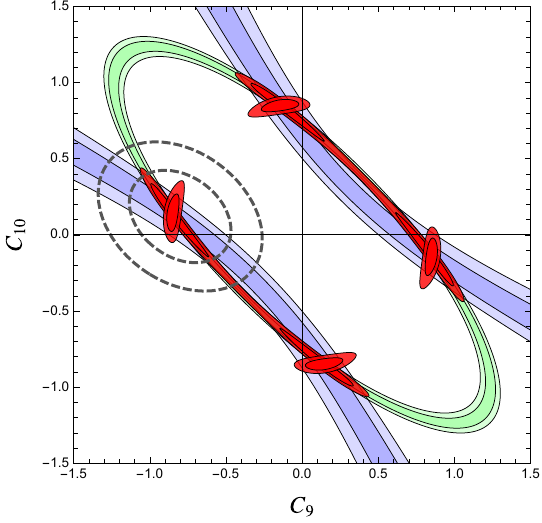}
\caption{Reach of a $10\,\text{TeV}$ muon collider in the $C_9$--$C_{10}$ plane, for $1\,\text{ab}^{-1}$ of data with unpolarised ({\it left}) and polarised ({\it right}) muon beams. Figures taken from \cite{Altmannshofer:2023uci}.\label{fig:mu}}
\end{figure}

The prospects for the scattering $\mu^+\mu^-\to bs$ at a $10\,\text{TeV}$ muon collider have been considered in~\cite{Altmannshofer:2023uci}. First, the measured cross-section determines the overall size of the NP effect (green circle in Fig.~\ref{fig:mu}). Making use of the forward-backward asymmetry (blue bands) provides additional information on the NP operator structure and helps to distinguish between vector and axialvector contributions. The achieved accuracy can be further increased by repeating the measurement with polarised muon beams, see the right panel of Fig.\ \ref{fig:mu}. It is important to note that the high-energy scattering $\mu^+\mu^-\to bs$ is insensitive to potential long-distance QCD effects.

Depending on the concrete realisation of NP behind the $b\to s\mu^+\mu^-$ anomaly and its flavour structure, the actual NP scale may be significantly lower than the SMEFT estimate which would greatly enhance the prospects of collider searches. Potential tree-level NP candidates are a $Z'$ gauge boson, or a scalar or vector leptoquark \cite{Capdevila:2023yhq}. In both cases lepton universality requires couplings also to electrons. This in turn leads to stringent LEP constraints on the $Z'$ solution, while for the leptoquark scenario two new particles are required coupling to either muons or electrons in order to comply with stringent limits on lepton flavour violating decays.

Alternatively, loop-induced NP in $b\to s\mu^+\mu^-$ would lower the relevant scale to the TeV range. In addition, it offers the intriguing possibility to link the anomalies in $b\to s\mu^+\mu^-$  and $\mathcal{R}(D^{(*)})$ to a common origin. For example, a vector leptoquark behind $\mathcal{R}(D^{(*)})$ can create a large contribution to $C_9$ via a $\tau$-lepton loop \cite{Bobeth:2011st,Crivellin:2018yvo,Aebischer:2022oqe} This scenario predicts large NP effects in $b\to s\tau^+\tau^-$ decays testable in $B$ experiments. Also a charged Higgs boson contributing to $\mathcal{R}(D^{(*)})$ can enter the relevant $b\to s\mu^+\mu^-$ Wilson coefficients through penguin and box diagrams \cite{Kumar:2022rcf}. This option can be tested in $t\tau^+\tau^-$ final states at the (HL-)LHC \cite{Iguro:2023jju}.

\section{Outlook}

The flavour anomalies discussed above are among the strongest hints for the presence of NP, with the appealing possibility of a common origin. In order to move forward on these (or potentially other future) anomalies in $B$ physics, we need an intertwined effort from both experiment and theory.

On the experimental side, future (even more) precise measurements of the relevant $B$ meson decays as well as better data on related decays of heavier $b$ hadrons (such as $\Lambda_b$, $B_c$) will shed light on the size and structure of the underlying flavour transition. At the same time, complementary information from NP searches and precision tests at high-energy hadron and lepton colliders will be indispensable to fully unravel the nature of the underlying physics.

From the theory side, an improved understanding of non-perturbative QCD effects entering the relevant decays will be needed to fully exploit the achieved experimental precision. In addition, the theory community will have to provide guidance for the experimental
community to identify the most promising observables. And last but certainly not least, 
we have to stay open-minded regarding the physics interpretation of the anomalies.

\paragraph{Acknowledgements} 

I would like to thank the organisers of the Corfu Summer Institute for inviting me to this pleasant and inspiring workshop.
Some of the research summarised here is supported by the Deutsche Forschungsgemeinschaft (DFG, German Research Foundation) under grant 396021762 -- TRR 257.

\end{document}